\documentclass{elsart}

\usepackage{amssymb}

\begin{document}

\begin{frontmatter}



\title{Fundamental Parameters of Low Mass X-ray Binaries II: X-Ray Persistent
Systems}


\author[IAC]{Jorge Casares,}
\author[SAAO]{Phil Charles}
\ead{jcv@iac.es; pac@saao.ac.za}

\address[IAC]{Instituto de Astrof\'\i{}sica de Canarias, 38200 -- La Laguna, 
Tenerife, Spain}
\address[SAAO]{South African Astronomical Observatory \& University of Southampton, UK}

\begin{abstract}
The determination of fundamental parameters in X-ray luminous (persistent)
X-ray binaries has been classically hampered by the large optical
luminosity of the accretion disc. New methods, based on irradiation of the
donor star and burst oscillations, provide the opportunity to derive 
dynamical information and mass constraints in many persistent systems for the 
first time. These techniques are here reviewed and the latest results 
presented. 
\end{abstract}

\begin{keyword}
binaries: close - X-rays: binaries,- stars: neutron
\PACS 
\end{keyword}
\end{frontmatter}

% main text
\section{Introduction}
\label{}
Sco X-1 was the first Galactic X-ray source discovered by X-ray
satellites and its powerful X-ray luminosity explained through mass
accretion into a degenerate star \cite{shk67}. Sco X-1 became the
prototype of a new class of objects, the low mass X-ray binaries
(hereafter LMXBs), where late-type, Roche lobe-filling stars transfer
matter onto compact objects \cite{van95}. Their orbital periods are
very compact, clustering at 3-6hr, and they mostly contain accreting
neutron stars \cite{liu01}. This is demonstrated by the exhibition of
Type I X-ray bursts (\cite{strob03}) and, in a few exceptional cases
(X1822-371 \cite{jonker01}, X1626-67 \cite{middle81}) X-ray
pulses. There are about 150 X-ray luminous or 'persistent' LMXBs in
the Galaxy, 30 of which with confirmed optical counterparts \cite{liu01}. For
comparison, it is estimated that the Galaxy contains a population of
$\sim 10^3$ 'transient' LMXBs which only show X-ray activity
sporadically and mostly harbour black holes (see accompanying review
by Charles \& Casares).

\begin{figure}
\begin{center}
\begin{picture}(250,160)(50,30)
\put(0,0){\includegraphics{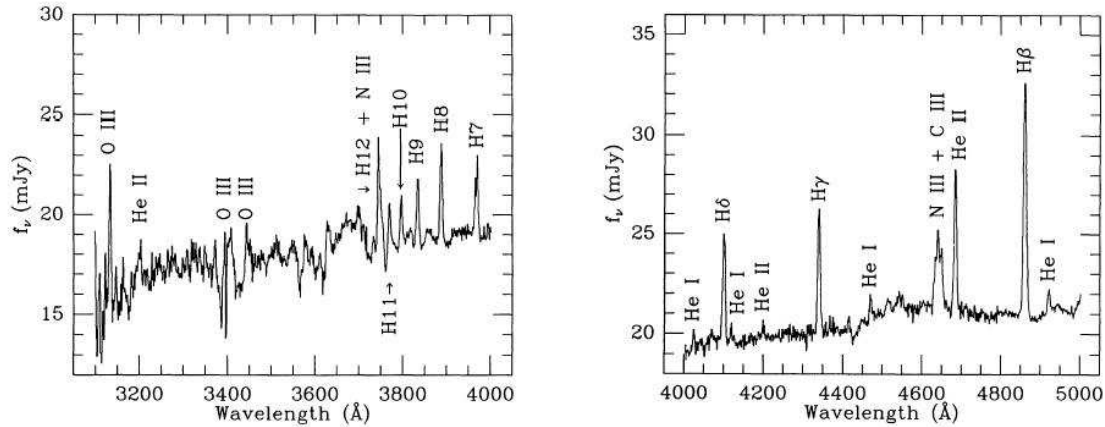}}
\noindent
\end{picture}
\end{center}
\caption{Optical spectrum of Sco X-1 (from \cite{scha89}) exhibits a flat energy 
distribution with superposed high excitation lines. Note the Bowen 
fluorescence CIII/NIII blend (right) and OIII cascade at 
$\lambda\lambda$3133, 3407, 3444 (left).}
\label{fig1}
\end{figure}

The reason for the different luminosity behaviour stems from the
interplay between the mass transfer rate from the donor star
$\dot{M_2}$ and irradiation effects. $\dot{M_2}$ determines the
temperature structure of the accretion flow and, if it is somewhere
below a critical temperature, $T_{crit}$ (that of H ionisation), then
instability cycles (outbursts) can be triggered (see
\cite{las01}). However, in persistent LMXBs, irradiation can keep the
outer disc hotter than $T_{crit}$ (even for low $\dot{M_2}\le 10^{-9}$
M$_{\odot}$ yr$^{-1}$), so that outburst cycles are suppressed and
discs appear persistently bright (\cite{van96}, \cite{king96}). They
display a large $L_X/L_{opt}$ and the optical and X-ray fluxes are
correlated, a strong indication that optical emission is caused by
reprocessing of higher energy photons by material in the vicinity of
the X-ray source.  Further evidence for X-ray reprocessing includes:

\begin{itemize} 

\item{} Statistical properties: dereddened optical colours follow the
distribution $(B-V)_0 =-0.09 \pm 0.17$ , $(U-B)_0 = -0.97 \pm 0.17$
\cite{van81}. This is consistent with F$_{\nu} \sim$ constant as
expected for the reprocessing of high energy photons.

\item{} Correlated X-ray/optical Type I bursts: the optical lags by a
few seconds, consistent with light travel times within the
binary. Optical profiles are smeared versions of the X-ray profiles,
indicating an extended reprocessing site (e.g. \cite{tru85}).

\item{} Presence of high excitation lines: e.g. HeII $\lambda$4686, 
CIII/NIII $\lambda\lambda$4640-50, and their 
flux is well correlated with $L_X$. Bowen fluorescence was 
proposed to explain the enhanced NIII emission \cite{mcc75}, a 
mechanism subsequently confirmed by the detection of 
OIII cascades in Her X-1 \cite{mar78} and Sco X-1 \cite{scha89} (Fig. 1). 

\end{itemize}

However, X-ray irradiation has systematically plagued 
attempts to determine system parameters in persistent LMXBs. These rely on 
dynamical information from the companion star, which is 
typically  $> 10^3$ times fainter than the X-ray heated accretion disc at 
optical-IR wavelengths. Fortunately, there are methods 
which can exploit the effects of irradiation and X-ray variability.
Here we provide an overview of these recent advances in the
determination of fundamental parameters of persistent LMXBS.

\section{Optical light curves}

Soon after the discovery of the first optical counterparts it was
clear that constraining binary parameters in X-ray bright LMXBs was a
difficult challenge \cite{van95}. Even the determination of the most
fundamental binary parameter, $P_{orb}$, proved elusive due to the
lack of obvious photometric variability. This led Milgrom \cite{mil78}
to propose a scenario where the companion star was effectively
shadowed from the central X-ray source by a flared accretion disc. The
absence of eclipsing systems was hence a pure selection effect. Since
then, the gain in X-ray sensitivity and deeper surveys has presented
several examples of eclipsing LMXBs and others with regular
optical/X-ray modulations and dips which indicate the binary periods 
\cite{liu01}.

Optical lightcurves are quasi-sinusoidal with superposed erratic 
variations and flickering, probably in response to variable X-ray illumination. 
It is widely assumed that optical maxima are associated with 
the irradiated inner face of the companion i.e. orbital phase $\phi=
0.5$.  This is supported by the relative phasing of X-ray eclipses
($\phi= 0$) and dips ($\phi= 0.8$)
(e.g. \cite{motch87}). Interestingly, the amplitude {\it A} seems to
be correlated with inclination \cite{van88}, from which LMXBs can be
placed into three broad categories: (i) {\it eclipsing}, with {\it A}
$\sim$0.5-1.5 mag and $i\ge 80^{\circ}$ (e.g.  X1822-371, X0748-636,
X2129+47) (ii) {\it dippers}, with {\it A} $\sim$0.5 mag and $i\simeq
70-80^{\circ}$ (e.g. X1254-690, X1755-338, X1916-05) and (iii) {\it low
i}, with {\it A} $\le$0.5 mag and $i\le 70^{\circ}$
(e.g. X1636-536, X1735-444, Sco X-1).\footnote{However, lightcurves of
the transient LMXB XTE J2123-058, obtained throughout the outburst cycle, show
a factor of 2 increase in optical amplitude when $L_X$
drops by one order of magnitude (see \cite{zur00},
\cite{sha03}). Therefore, a straight correlation between {\it A}
and {\it i} should be treated with caution.} 

Tighter constraints on $i$ can be set by detailed modelling of
optical/X-ray eclipses and light curves of accretion disc coronae
(ADC) sources \cite{white82}. In particular, simultaneous fits to
EXOSAT X-ray/optical light curves in X1822-371 have led to the most
accurate determination of the disc geometry and $i$ ($83^{\circ} \pm
2^{\circ}$) in an LMXB \cite{hell89}, \cite{hein01}. The model
includes irradiation, shadowing and obscuration by the disc, donor
star, bulge and ADC structures (see Fig. 2). 
   
\begin{figure}
\begin{center}
\begin{picture}(250,140)(50,30)
\put(0,0){\includegraphics{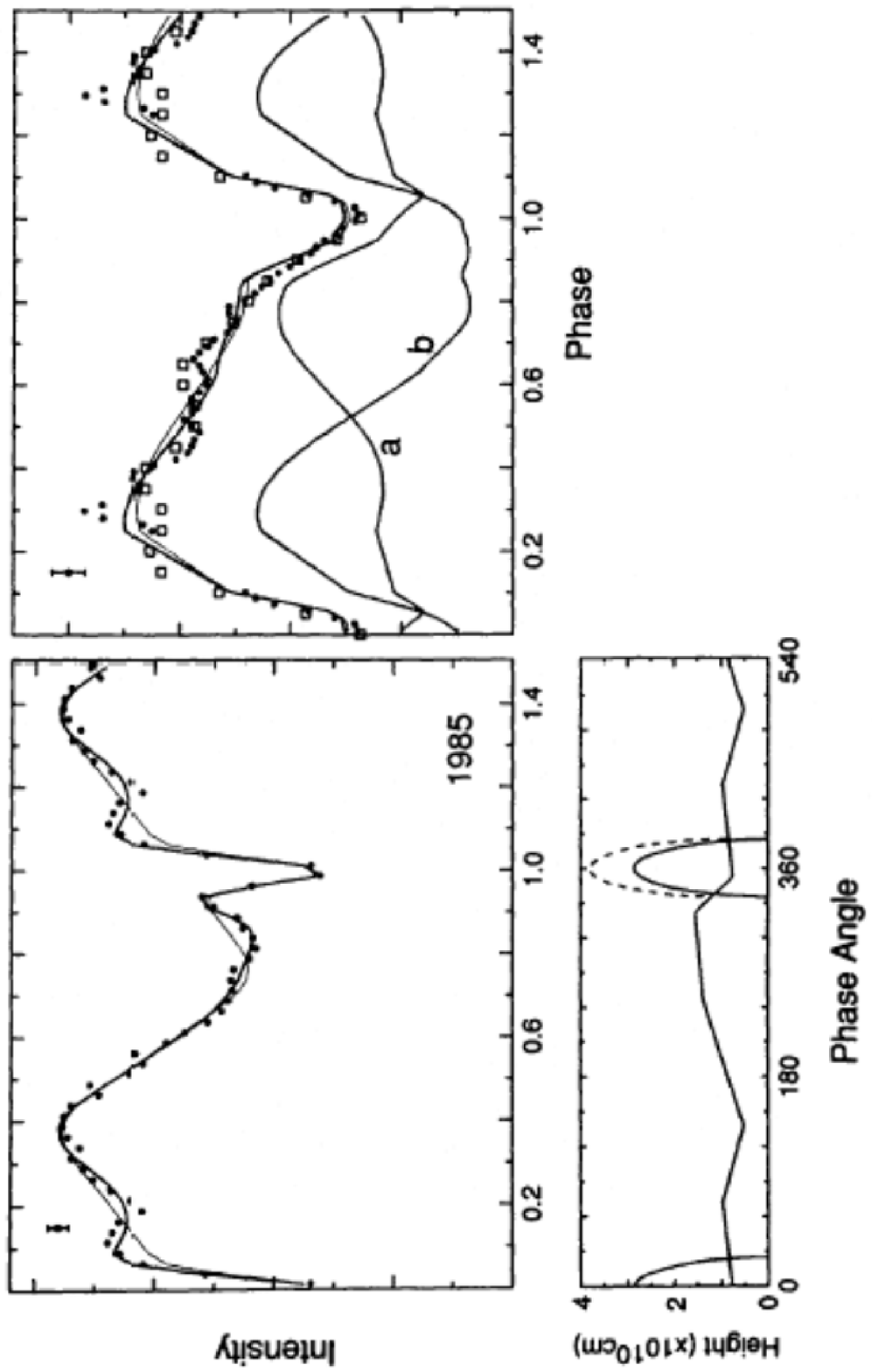}}
\put(0,0){\includegraphics{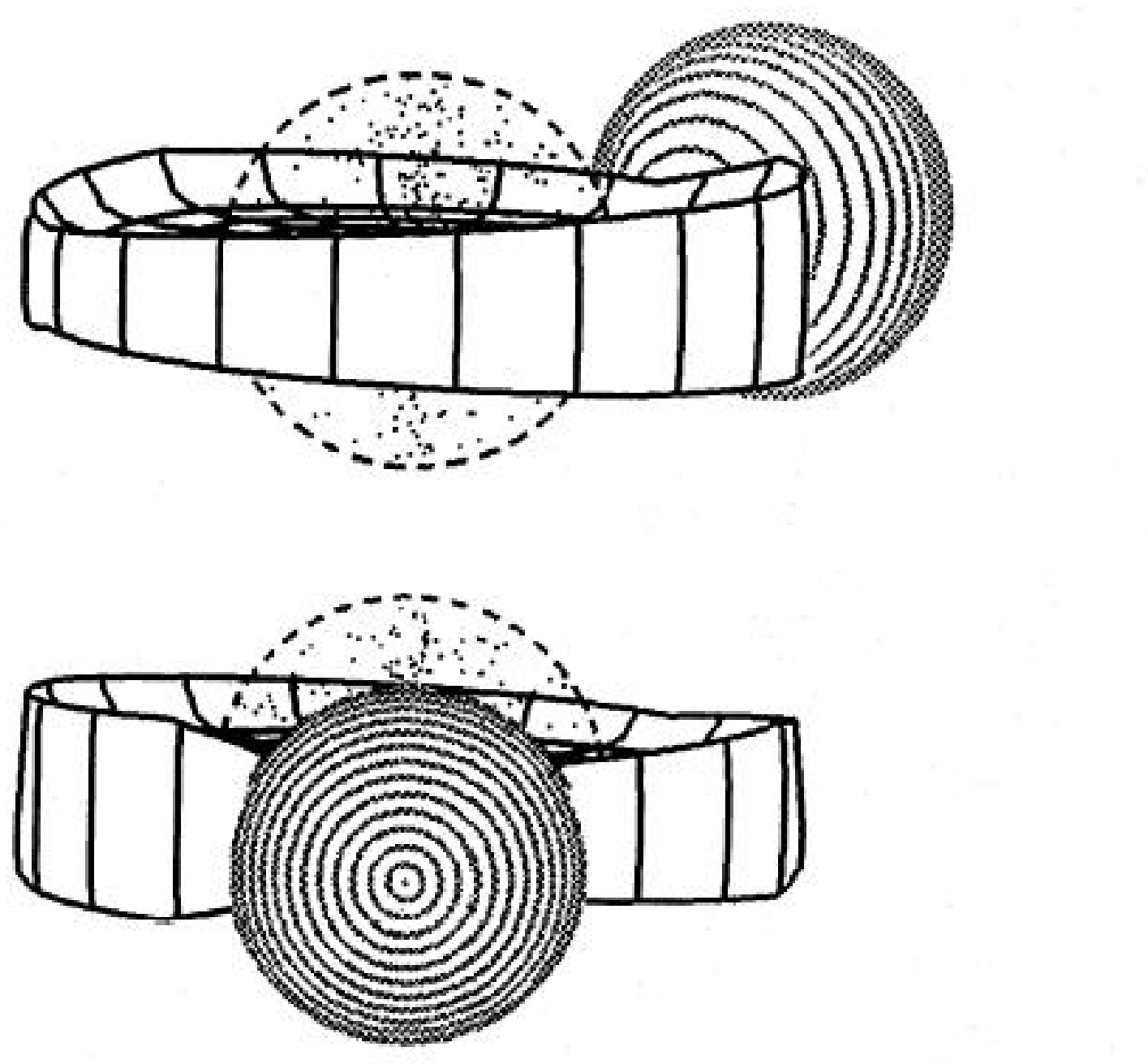}}\noindent
\end{picture}
\end{center}
\caption{EXOSAT and optical light curves (left) of X1822-371 and 
best-fit model, with the disc and ADC shown schematically (right), from 
\cite{hell89}.}
\label{fig2}
\end{figure}

\section{Dynamical Information}
 
With the exception of a few long period ($P> 1$ d) systems with
evolved donors (e.g. Cyg X-2 \cite{casa98}, X0921-630 \cite{sha04}),
spectroscopic features of companion stars in persistent LMXBs are
totally veiled by the accretion disc continuum.
Attempts to derive dynamical information of the compact star using 
emission lines (mainly Balmer or HeII $\lambda$4686) have proven unreliable 
because the emission lines are very broad and show a 
complex, variable, multi-component structure. The bulk of the emission
tends to be dominated by the disc bulge with superior   
conjunction at orbital phase $\sim$0.75\footnote{As determined by photometric
ephemerides from eclipses or lightcurve minima.} e.g. 
X1636-536, X1735-444 \cite{aug98}. 
 
\subsection{Fluorescent emision from the companion}

New prospects for dynamical studies have been opened by the discovery of
high-excitation 
emission lines arising from the donor star in Sco X-1 \cite{stee02}.
The most prominent are in the core of the Bowen blend, namely the
triplets NIII $\lambda$4634-40 and CIII $\lambda$4647-50.
In particular, the NIII lines are powered by fluorescence resonance which
requires seed photons of HeII Ly-$\alpha$.
These narrow components move in phase with each other and are not resolved
(i.e. their FWHM is the instrumental resolution, 50 km s$^{-1}$), an 
indication that the reprocessing region is very localized
(Fig. 3). The extreme narrowness rules out the accretion flow or the hot
spot and points to the companion star as the reprocessing site.

\begin{figure}[ht]
\begin{center}
\begin{picture}(250,150)(50,30)
\put(0,0){\includegraphics{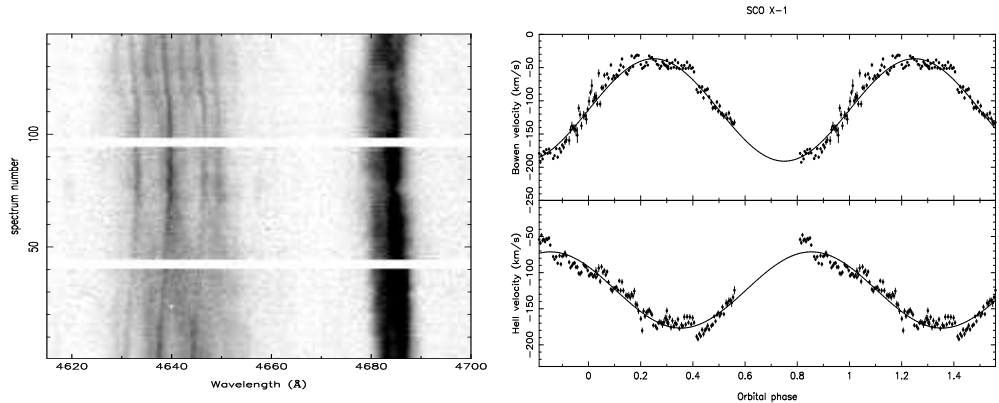}}
\put(0,0){\includegraphics{fig3b.eps}}
\noindent
\end{picture}
\end{center}
\caption{Trailed spectra (left) and radial velocity curves (right) of the
narrow CIII/NIII (top) and broad HeII $\lambda$4686 (bottom) emission lines 
in Sco X-1 (from \cite{stee02}).}
\label{fig3}
\end{figure}

The radial velocity curve of 
the Bowen lines (Fig.3) is in antiphase
with the HeII $\lambda$4686 wings, which approximately
trace the motion of the compact star. 
Furthermore, they are also in phase with the maximum of the photometric 
light curve,  ascribed to the irradiated face of the donor star \cite{aug92}.  
This work represents the first detection of the companion star in Sco X-1 
and has opened a new avenue for dynamical studies of 
luminous LMXBs. 

We currently know that fluorescence is not peculiar to Sco X-1, 
but is a general signature of active LMXBs. This is exemplified by recent work 
on X1822-371, the archetypal eclipsing ADC which also contains a 0.59s pulsar 
\cite{jonker01}. 
Both $i$ and the pulsar orbit are extremely well constrained 
and hence only the radial velocity curve of the companion star is  
needed for a full determination of the system parameters. 
The faintness of X1822-371, coupled with the high spectral resolution 
($\le70$ km s$^{-1}$) required to resolve the Bowen lines, prevents 
detection in individual spectra. However, exploiting 
Doppler Tomography~\cite{marsh01} we can 
combine all the information contained in individually phase-resolved 
spectra to reconstruct the emissivity distribution. Fig. 4 shows the Doppler maps 
of HeII $\lambda$4686 and NIII $\lambda$4640 in velocity space.  
Unlike HeII $\lambda$4686 (which shows the classic 
accretion disc ring) the NIII map displays a compact spot located at the 
expected velocity and phasing of the donor star, as predicted from the pulsar 
ephemeris. The centroid of the spot lies at 300 km s$^{-1}$ and establishes a 
lower limit to the velocity of the companion because it arises from the 
irradiated inner face. In order to derive the true velocity 
(and then accurate masses) one needs to model the displacement between the 
reprocessing site and the donor's center of mass as a function of the 
mass ratio {\it q}. The so-called {\it K-correction} depends on details of 
reprocessing physics and irradiation geometry, such as shielding effects by 
the disc (see \cite{teo05}). 

\begin{figure}
\begin{center}
\begin{picture}(250,140)(50,30)
\put(0,0){\includegraphics{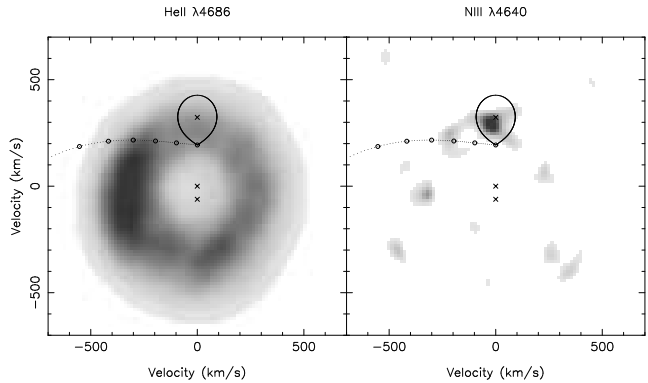}}
\noindent
\end{picture}
\end{center}
\caption{Doppler images of HeII $\lambda$4686 (left) and NIII
$\lambda$4640 (right) for X1822-371 (from \cite{casa03}). The Roche
lobe of the donor and gas stream trajectory are overplotted
for a 1.4 M$_{\odot}$ neutron star.}
\label{fig4}
\end{figure}

Follow-up campaigns, using 8m-class telescopes (e.g. VLT), has enabled
us to extend this analysis to fainter LMXBs. This has led to the first
detection of the donors in X1636-536, X1735-444 and GX 9+9  and
the determination of their orbital velocities, which lie in the range
200-300 km s$^{-1}$ (\cite{casa03}, \cite{casa04}).  In addition, this
technique has been applied to transient LMXBs in outburst, such as the
millisecond pulsar XTE J1814-338, Aql X-1 and the black hole candidate
GX339-4~\cite{hynes03}. In the latter case, the observations provided the first
dynamical proof that it is a black hole.

\subsection{Burst Oscillations}

Despite LMXBs having long been considered the progenitors of
millisecond pulsars, such pulsations have escaped detection until
recently.  This changed thanks to RXTE with the discovery of: (i)
persistent pulses in 5 transient LMXBs with $P_{spin}$ in the range
185-435 Hz, and (ii) nearly coherent oscillations during X-ray bursts
in 13 luminous LMXBs.  In SAX J1808-3658 \cite{chak03} and XTE
J1814-338 \cite{stro03} both burst oscillations and persistent pulses
were detected and with identical frequencies, confirming that burst
oscillations are indeed modulated with the neutron star spin. 
Furthermore, a smooth frequency drift in the oscillation could be 
observed during a {\it superburst} in X1636-536, caused by the doppler motion
of the neutron star \cite{stro02}.  Burst oscillations can therefore be used to
trace the neutron star orbit in persistent LMXBs and, in combination
with Bowen fluorescence, these luminous LMXBs can become double-lined
spectroscopic binaries.

\section{Echo-tomography}

Echo Tomography is a powerful technique which employs time delay between 
X-ray and UV/optical variability to map the reprocessing sites in a 
binary \cite{obrien02}. The optical lightcurve results from the convolution of 
the X-ray lightcurve with a {\it transfer function} representing the binary 
response to the illuminating flux. The transfer function contains a 
phase-dependent component, associated with the donor, which encodes 
information on the most fundamental parameters, namely $i$, 
$a$ and $q$. 
Succesful Echo-tomography experiments have been performed in 
several X-ray active LMXBs using X-ray and broad-band UV/optical lightcurves. 
However, the results indicate that the reprocessing flux is mostly dominated by 
the accretion disc with no orbital phase dependency (e.g. \cite{van90}, 
\cite{hynes05}). Exploiting emission-line reprocessing rather than broad-band
photometry has two potential benefits: a) it amplifies the response of
the donor's contribution by suppressing most of the background
continuum light (dominated by the disc); b) since the
emission line reprocessing time is instantaneous, the response is
sharper (i.e. only smeared by geometry). Recent results on Sco X-1, using 
narrow-band filters centered on the Bowen blend + HeII $\lambda$4686 region,    
simultaneously with RXTE, have shown evidence for delayed echoes associated 
with the donor (see Mu\~noz-Darias et al., these 
proceedings). 
   
\section{Conclusions}

The study of Bowen fluorescence in X-ray active LMXBs is producing
significant progress in the determination of fundamental system
parameters.  This is possible because of: i) dynamical information
obtained by detecting the donors through high-resolution spectroscopy
of the Bowen blend; ii) echo-mapping reprocessing sites through
simultaneous Bowen-line/X-ray lightcurves.  These techniques, together
with results from burst oscillations, will likely provide the first
accurate neutron star masses in luminous LMXBs in the near
future. High-speed and high-resolution instruments in new generation
large telescopes (such as OSIRIS at GTC and PFIS at SALT) will play a
crucial role in this goal.

\end{document}